# An Overview of Recent Developments in 'Big' Boolean Equations

**Ali Muhammad Ali Rushdi**

*Department of Electrical and Computer Engineering, Faculty of Engineering,
King Abdulaziz University P. O. Box 80200, Jeddah 21589, Saudi Arabia*
{arushdi@ieee.org; arushdi@kau.edu.sa}

**Abstract.** A task frequently encountered in digital circuit design is the solution of a two-valued Boolean equation of the form $h(X, Y, Z) = 1$, where $h: B_2^{k+m+n} \to B_2$ and $X, Y$, and $Z$ are binary vectors of lengths $k$, $m$, and $n$, representing inputs, intermediary values, and outputs, respectively. The *resultant* of the *suppression* of the variables $Y$ from this equation could be written in the form $g(X, Z) = 1$ where $g: B_2^{k+n} \to B_2$. Typically, one needs to solve for $Z$ in terms of $X$, and hence it is unavoidable to resort to 'big' Boolean algebras which are finite (atomic) Boolean algebras larger than the two-valued Boolean algebra. Other situations which necessitate 'big' Boolean-equation-solving are the so-called 'elementary problems of digital circuit design,' which entail five vector-Boolean quantities $X, Y, Z, s$ and $t$ that belong to $B_2^k, B_2^m, B_2^n, B_2^l$, and $B_2^l$, respectively. Consider a 'parent' combinational network $C$ of two (vectorial) inputs $X$ and $Y$ and a vectorial output $t(X, Y)$, and assume that network $C$ consists of two subnetworks $A$ and $B$, where subnetwork $A$ has the single (vectorial) input $X$ and the (vectorial) output $Z(X)$, while network $B$ has the two vectorial inputs $Z(X)$ and $Y$ and the (vectorial) output $s(Z, Y)$, which is exactly the same as the (vectorial) output $t(X, Y)$ of network $C$. The above arrangement involves three vectorial Boolean functions, namely $Z(X)$, $s(Z, Y)$ and $t(X, Y)$. Three problems arise when one utilizes the information that $s(Z(X), Y)$ and $t(X, Y)$ are equal, together with knowledge of two of the three functions $Z(X)$, $s(Z, Y)$ and $t(X, Y)$ in order to deduce the third function. Two of these problems are natural examples why 'big' Boolean-equation-solving is warranted. Methods of solving 'big' Boolean equations can be broadly classified as algebraic, tabular, numerical and map methods. The most prominent among these classes are the algebraic and map methods. This paper surveys and compares these two types of methods as regards simplicity, efficiency, and usability. The paper identifies the main types of solutions of Boolean equations as subsumptive general solutions, parametric general solutions and particular solutions. In a subsumptive general solution, each of the variables is expressed as an interval based on successive conjunctive or disjunctive eliminants of the original function. In a parametric general solution each of the variables is expressed via arbitrary parameters which are freely chosen elements of the underlying Boolean algebra. A particular solution is an assignment from the underlying Boolean algebra to every pertinent variable that makes the Boolean equation an identity. The paper offers a tutorial exposition, review, and comparison of the three types of solutions by way of an illustrative example. This example is also viewed from the perspective of a recently developed technique that relies on atomic decomposition. Map techniques are demonstrated to be at least competitive with (and occasionally superior to) algebraic techniques, since they have a better control on the minimality of the pertinent function representations, and hence are more capable of producing more compact solutions.

# 1. Introduction

A prominent "misnomer" in mathematical and engineering circles is the term 'Boolean algebra'. This term is widely used to refer to switching algebra, which is just one particular case of a finite 'Boolean algebra' that has 0 generators, 1 atom and two elements belonging to $B_2 = \{0, 1\}$. The term 'Boolean algebra' refers to an algebra of a finite or infinite cardinality (Halmos and Givant, 2009). The term 'finite Boolean algebra' covers, in fact, a countably infinite number of atomic algebras described by natural numbers $n$ ($n \geq 0$), such that an algebra has n generators, N atoms ($2^{n-1} < N \leq 2^n$), and $2^N$ elements. The inadvertent use of the general term Boolean algebra to refer to its particular case of a switching algebra $B_2$ leads to many problems, fallacies and misconceptions, since the switching algebra is much simpler than a generalized finite (atomic) Boolean algebra. Therefore, many authors (Ahmad and Rushdi, 2018; Brown, 1990; Rushdi and Ahmad, 2017; 2018; Rushdi and Amashah, 2012; Rushdi and Zagzoog, 2019) started to label a finite Boolean algebra other than $B_2$ (i.e., one with n generators ($n > 0$)) as a "big" Boolean algebra. Figure 1 Visualizes the carriers (complemented distributive lattices) of the two 'big' Boolean algebras $B_8$ and $B_{16}$, demonstrating 'partial ordering' among their elements. Here, the hypercube $B_{16}$ is the free Boolean algebra $FB(a, b)$. It collapses to a cube $B_8$ when one of its four atoms (here $ab$) is nullified. Both algebras need (at least) two generators $a$ and $b$.

The study of Boolean equations in 'big' Boolean algebras spanned the past two centuries with fundamental contributions from prominent scholars such as Boole, Schröder, Whitehead, Poretski, Löwenheim, Rudeanu, and Brown. This fact asserts that we *stand on the shoulders of giants*, indeed. The importance of 'big' Boolean-equation solving can hardly be overestimated. It permeates many subareas of digital or logical design, such as decomposition of Boolean functions, fault diagnosis and hazard-free synthesis of digital circuits, construction of binary codes, flip-flop excitation, and the design of sequential circuits. Other areas of modern science to which 'big' Boolean-equation solving has been applied include biology, grammars, chemistry, law, medicine, spectroscopy, and graph theory (Brown, 1990).

Methods of solving 'big' Boolean equations can be broadly classified as algebraic, tabular, numerical and map methods. The most prominent among these classes are the algebraic and map methods. This paper surveys and compares these two types of methods as regards simplicity, efficiency, and usability. The paper identifies the main types of solutions of 'big' Boolean equations (Rudeanu, 1974; 2010; Brown; 1970; 1974; 1990; Rushdi, 2001; 2004; Baneres *et al.*, 2009; Rushdi and Ba-Rukab, 2003) as subsumptive general solutions, parametric general solutions and particular solutions. In a subsumptive general solution, each of the variables is expressed as an interval based on successive conjunctive or disjunctive eliminants of the original function. In a parametric general solution each of the variables is expressed via arbitrary parameters which are typically chosen freely as elements of the underlying Boolean algebra. A particular solution is an assignment from the underlying Boolean algebra to every pertinent variable that makes the Boolean equation an identity. This paper offers a tutorial exposition, review, and comparison of the three types of solutions by way of an illustrative example. This example is also viewed from the perspective of a recently developed technique that relies on the atomic decomposition of pertinent functions and variables. Map techniques are demonstrated to be at least competitive with (and occasionally superior to) algebraic techniques, since they have a better control on the minimality of the pertinent function representations, and hence are more capable of producing more compact solutions.

The organization of the remainder of this paper is as follows. Section 2 introduces a variety of specific situations in which 'big' Boolean-equation solving is unavoidable. Section 3 covers the mathematical details of various solution methods and related issues such as how to unify given digital specifications (such as the ones in the form $s = t$) into a single Boolean equation and the method of suppression of variables. In particular, Section 3 derives parametric and subsumptive general solutions of an arbitrary Boolean equation, and explains how these can be recast into a very compact listing of all particular solutions, which enables one to easily locate particular solutions of certain desirable features. Section 4 compares algebraic and map methods for solving Boolean equations. Section 5 presents a detailed illustrative example in which a general parametric solution is obtained via three methods, which are (a) a direct compact solution, (b) a compact solution obtained indirectly via a subsumptive solution, and (c) a direct permutative additive solution leading to compact listing of all particular solutions. This permutative additive solution is a forerunner of solutions obtained via the atomic decomposition of pertinent functions and variables. Section 6 concludes the paper.

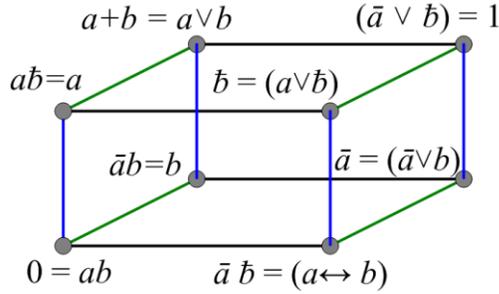
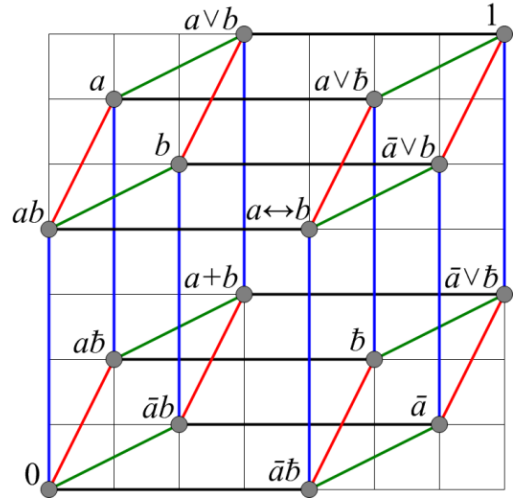

The lattice of $B_8$, obtained when $B_{16}$ is collapsed under the condition of nullifying one of the four atoms (ab = 0).

Hypercube lattice indicating the partial ordering among the 16 elements of $B_{16}$, an example of which is
$0 \leq \bar{a}\bar{b} \leq \bar{a} \leq \bar{a} \vee \bar{b} \leq 1$.

Fig. 1. Visualization of two 'big' Boolean algebras $B_8$ and $B_{16}$. Here, the hypercube $B_{16} = FB(a, b)$ collapses to a cube $B_8$ when one of its four atoms (here $ab$) is nullified. Both algebras need (at least) two generators $a$ and $b$.

## 2. The Need for 'Big' Boolean-Equation Solving

We now demonstrate a variety of specific situations in which 'big' Boolean-equation solving is unavoidable. The simplest such situation involves solving an 'inverse problem of logic' in which knowledge of the vectorial function $Z(X)$ is utilized to produce its inverse vectorial function $X(Z)$ (Jevons, 1872; 1874; Brown, 1974; 1975b). When $X$ and $Z$ are binary vectors of lengths $k$ and $n$, representing inputs and outputs, respectively, of a digital circuit, then the aforementioned problem amounts to deducing the inverse circuit function $X(Z)$ through knowledge of the forward one $Z(X)$. Knowledge of the function $Z(X)$ might be reformulated as an equation of the form $g(X, Z) = 1$ where $g: B_2^{k+n} \to B_2$. This equation is then solved for $X$, with its elements belonging to the free Boolean algebra $FB(Z)$, i.e., the Boolean algebra whose elements constitute all the $2^{2^n}$ switching functions of the $n$ variables $Z_1, Z_2, \ldots$ and $Z_n$.

A more involved situation concerns a task frequently encountered in digital circuit design, in which circuit specifications can be (and typically are) reduced to a single two-valued Boolean equation of the form $h(X, Y, Z) = 1$, where $h: B_2^{k+m+n} \to B_2$ and $X$ and $Z$ are as before, while $Y$ is a binary vector of length $m$, representing intermediary values (typically inaccessible and hence uncontrollable and unobservable). The above equation might be used to obtain all solutions of $Z$ over $FB(X,Y)$, and then only solutions of $Z$ that depend solely on $X$ are retained. These solutions, which are independent of $Y$, might be directly obtained through the derivation of the *resultant* of the *suppression* of the variables $Y$ from the original equation, which could be cast in the form $g(X,Z) = 1$ where $g: B_2^{k+n} \to B_2$. The required solutions are obtained simply from $g(X,Z) = 1$ being solved for $Z$ over $FB(X)$ (Brown, 2011).

More sophisticated situations of digital circuit design can be described by the general layout presented in Fig. 2. This layout was first suggested, without the shady sneaky path, by Ledley (1959; 1960), and later used by Bell (1968), and Brown (1975a). Ledley's layout involves a combinational logic circuit C that is comprised of two sub-networks A and B. The final output can be viewed as output $s(Y,Z)$ of sub-network B or output $t(X,Y)$ of the parent network C. Here, the five quantities $X, Y, Z, s$ and $t$ are binary or Boolean vectors of lengths $k, m, n, l$, and $l$ representing basic inputs, side inputs, intermediary values, output for subnetwork B, and output for the overall network C, respectively. This means that the quantities $X, Y, Z, s$ and $t$ belong to $B_2^k, B_2^m, B_2^n, B_2^l$, and $B_2^l$, respectively, where $B_2$ stands for the two-valued Boolean algebra $B_2$. Consequently, three 'elementary problems of digital circuit design' emerge (Ledley, 1959; 1960; Rushdi & Ahmad, 2017), namely

Type-1 problem: Given $Z(X)$ and $s(Y,Z)$, find $t(X,Y)$.
Type-2 problem: Given $Z(X)$ and $t(X,Y)$ find $s(Y,Z)$.
Type-3 problem: Given $s(Y,Z)$ as well as $t(X,Y)$, find $Z(X)$.

Brown (1975a) slightly altered the structure of Ledley's network by adding the shady sneaky path in Fig. 2, and, as a result, replacing $s(Y,Z)$ by $s(X,Y,Z)$. Sub-network B is called an 'honest translator' or a 'sneaky translator' depending on whether the shaded path in Fig. 1 can be omitted or it is needed (or used).

While Type-1 problems can be trivially solved by direct substitution, Type-2 and Type-3 problems might be handled via 'big' Boolean-equation solving. Addressing Type-3 problems relies solely on the methodology of equation solving in 'big' Boolean algebras (Rushdi and Ahmad, 2017), but handling generalized Type-2 problems is better achieved via other techniques (Brown, 1975a; Rushdi, 2018: 2019).

A prominent potential area for 'big' Boolean algebra application is that of cryptography. Ahmad and Rushdi (2018) have recently proposed a scheme that can encrypt a message by using a 'big' Boolean function, which produces an equation that cannot be solved by conventional satisfiability (SAT) Solvers and leads to a dramatic increase in the search space in the worst case. Logical cryptanalysis shows that the proposed scheme is very hard to break, indeed. It seems that the adversary cannot reduce or prune the search space (except for shortening the task needed at every node), and is forced to traverse the whole search space. In fact, the adversary might arrive at several candidate solutions, and then has to search for clues as to which of them is the correct solution. The adversary has to confront not only the excessive difficulty for finding the solution of a 'big' Boolean equation, but also the ambiguity of non-uniqueness of this solution. The aforementioned scheme brought about a revival in the problem of Boolean curve fitting (BCF), or Boolean interpolation, which has been viewed as a pure mathematical curiosity for one full century. A unified solution of the BCF problem is derived by solving a system of Boolean equations over a 'big' Boolean algebra (Rudeanu, 1974; Rushdi and Balamesh, 2019; Balamesh and Rushdi, 2022).

### 3. Mathematical Derivations for Various Solution Types
#### 3.1. Unifying Specifications into a Single Boolean Equation

A remarkable feature of any Boolean algebra is that a system of several equations is equivalent to a single equation. For example, consider the digital system of Fig. 2, which is specified by a system of $l$ scalar Boolean equations of the vector form

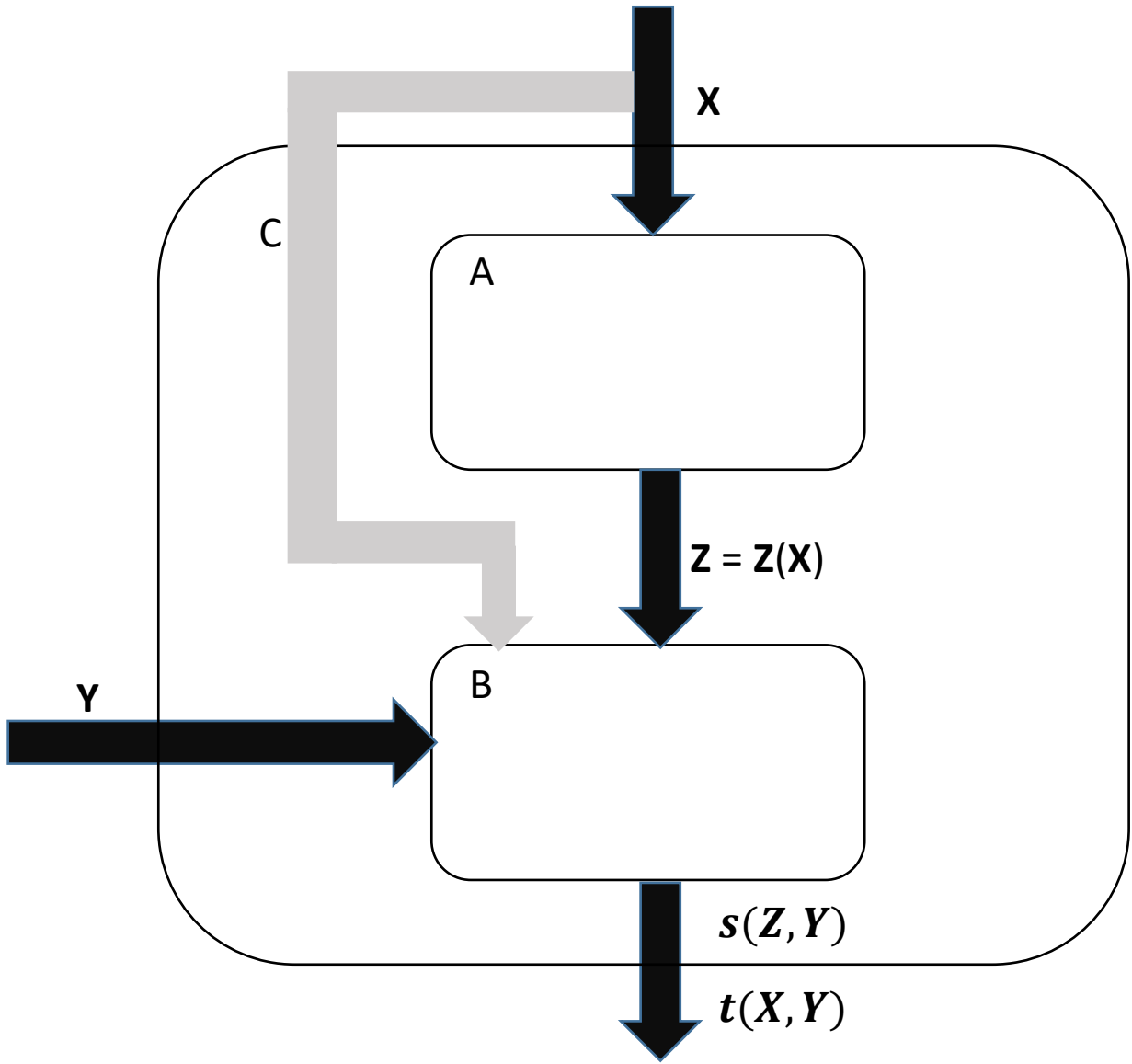

Fig. 2. Layout of generalized problems of digital circuit design (If the sneaky (dotted) path is needed, the function $s(Z.Y)$ is replaced by $s(X,Z,Y)$).

$$\boldsymbol{s = t}, \qquad (1a)$$

or of the equivalent scalar form

$$s_i = t_i, \quad 1 \le i \le l, \qquad (1b)$$

where $s_i = s_i(\mathbf{X}, \mathbf{Y}, \mathbf{Z})$ and $t_i = t_i(\mathbf{X}, \mathbf{Y}, \mathbf{Z})$. The system (1) of $l$ scalar equations reduces to a single Boolean equation of the form

$$h(\boldsymbol{X}, \boldsymbol{Y}, \boldsymbol{Z}) = 1, \qquad (2a)$$

where

$$h(X, Y, Z) \equiv \wedge_{i=1}^{l} (s_i \odot t_i), \tag{2b}$$

or of the form

$$r(X, Y, Z) = 0, \tag{3a}$$

where

$$r(X, Y, Z) = \bar{h}(X, Y, Z) \equiv \vee_{i=1}^{l} (s_i \oplus t_i). \tag{3b}$$

The symbols $\wedge$, $\vee$, $\oplus$, and $\odot$ in Equations (2b) and (3b) depict the AND operator, the OR operator, the XOR (Exclusive-OR) operator and the XNOR (coincidence or equivalence) operator, respectively. The relation between (2b) and 3(b) is just an expression of the two De' Morgan's laws. Note that the AND and OR operators are dual ones, while the XNOR and XOR operators are both complementary and dual ones.

### 3.2. Suppression of Variables

Brown (2011) proved that the *resultant of suppression* of the variables $Y$ from the Boolean equation (3a) (called the parent equation) is the derived Boolean equation

$$f(X, Z) = 0, \tag{4a}$$

where

$$f(X, Z) \equiv \vee_{A \in \{0,1\}^m} r(X, A, Z). \tag{4b}$$

and that the solutions of the derived equation (4a) are exactly those of the parent equation (3a) that do not involve the suppressed variables $Y$.

We will herein utilize the *dual* of the above result, namely that if we use (2a) instead of (3a) as a parent equation, then the resultant of suppression of the variables $Y$ is now the complementary derived Boolean equation

$$g(X, Z) = 1, \tag{5a}$$

where

$$g(X, Z) \equiv \wedge_{A \in \{0,1\}^m} h(X, A, Z). \tag{5b}$$

and the solutions of the derived equation (5a) are exactly those of the parent equation (2a) that do not involve the suppressed variables $Y$.

### 3.3. Derivation of Parametric Solutions

We seek solutions of a Boolean equation of the form (5a), where $g(X, Z): B_2^{k+n} \to B_2$, is a two-valued Boolean function of $k$ two-valued variables $X = [X_1 \ X_2 \ ... \ X_k]^T$ and $n$ two-valued variables $Z = [Z_1 \ Z_2 \ ... \ Z_n]^T$. However, we do not need a listing of binary solutions for $X$ and $Z$, but instead we want to express $Z$ in terms of $X$. This means that we are seeking the solution of a 'big' Boolean equation. We view $g(X, Z)$ as $g(X; Z)$ or simply $g(Z)$ and rewrite (5a) as

$$g(Z) = 1, \tag{6}$$

where $g(\mathbf{Z}): B_{2^K}^n \to B_{2^K}$, and $B_{2^K}$ is the free Boolean algebra $FB(X_1, X_2 \ldots \ldots X_k)$ with $K = 2^k$ atoms and $2^K = 2^{2^k}$ elements. Now we express $g(\mathbf{Z})$ by its Minterm Canonical Form (MCF) (Brown, 1990)

$$g(\mathbf{Z}) \equiv \vee_{\mathbf{A} \in \{0,1\}^n} g(\mathbf{A}) \, \mathbf{Z}^{\mathbf{A}}. \tag{7}$$

For $\mathbf{Z} = [\, Z_1 \, Z_2 \ldots Z_n \,]^T \in B_{2^K}^n$, $\mathbf{A} = [\, a_1 \, a_2 \ldots a_n \,]^T \in \{0,1\}^n$, the symbol $\mathbf{Z}^{\mathbf{A}}$ is defined as

$$\mathbf{Z}^{\mathbf{A}} = Z_1^{a_1} \, Z_2^{a_2} \ldots Z_n^{a_n}, \tag{8}$$

where $Z_i^{a_i}$ takes the value $\bar{Z}_i$ (complemented literal) if $a_i = 0$, and takes the value $Z_i$ (uncomplemented literal) if $a_i = 1$. For $\mathbf{A} \in \{0,1\}^n$, the symbol $\mathbf{Z}^{\mathbf{A}}$ spans the minterms of $\mathbf{Z}$, which are the $2^n$ elementary or primitive products

$$\bar{Z}_1 \bar{Z}_2 \ldots \bar{Z}_{n-1} \bar{Z}_n, \quad \bar{Z}_1 \bar{Z}_2 \ldots \bar{Z}_{n-1} Z_n, \quad \ldots, \quad Z_1 Z_2 \ldots Z_{n-1} Z_n. \tag{9}$$

The constant values $g(\mathbf{A})$ in equation (7) are elements of $B_{2^K}$ called the discriminants of $g(\mathbf{Z})$. These discriminants are the entries of the natural map of $g(\mathbf{Z})$ which has an input domain $\{0,1\}^n \subseteq B_{2^K}^n$. The Boolean algebra $B_{2^K} = FB(X_1, X_2 \ldots \ldots X_k)$, has generators $X_i$ ($1 \leq i \leq k$) which look like variables (In fact, they were originally our input variables before we changed their roles to generators). Therefore, we can accept the name assigned (for historical reasons) to the natural map of $g(\mathbf{Z})$, namely, the name of the Variable-Entered Karnaugh Map (VEKM). We now observe that the minterms of $\mathbf{X}$, which are the $2^k = K$ elementary or primitive products

$$\bar{X}_1 \bar{X}_2 \ldots \bar{X}_{k-1} \bar{X}_k, \quad \bar{X}_1 \bar{X}_2 \ldots \bar{X}_{k-1} X_k, \quad \ldots, \quad X_1 X_2 \ldots X_{k-1} X_k, \tag{10}$$

are exactly the atoms of the underlying Boolean algebra. For convenience, we call these atoms $T_i$ ($0 \leq i \leq (K-1)$), and hence $g(\mathbf{A})$ can be written as

$$g(\mathbf{A}) = \vee_{i=0}^{K-1} (e_i(\mathbf{A}) \wedge T_i), \tag{11}$$

where we use the symbol $e_i(\mathbf{A})$ to denote an indicator of the event that atom $T_i$ appears in the expression of $g(\mathbf{A})$, i.e.,

$$e_i(\mathbf{A}) = \begin{cases} 1, & \text{if } T_i \to g(\mathbf{A}) \\ 0, & \text{otherwise} \end{cases} = g(\mathbf{A})/T_i, \tag{12}$$

where the symbol $(r \,/\, s) = (r)_{s=1}$ denotes the Boolean quotient of $r$ by $s$ (Brown, 1990). Equation (12) means that $e_i(\mathbf{A})$ indicates whether atom $T_i$ appears in the cell $\mathbf{A}$ of the natural map for $g(\mathbf{Z})$. Now, we define $n_i$ ($0 \leq n_i \leq 2^n$) as the total number of actual appearances of $T_i$ in the expression (11) for $g(\mathbf{A})$, i.e.,

$$n_i = \sum_{\mathbf{A} \in \{0,1\}^n} e_i(\mathbf{A}). \tag{13}$$

The total number $N_{unconditional}$ of unconditional particular solutions of (1a) over $B_{2^K}$ (as it is) is given by

$$N_{unconditional} = \prod_{i=0}^{K-1} n_i. \tag{14}$$

This number is zero if some $n_i = 0$, *i.e.*, if an atom $T_i$ never makes its way to any expression $g(\mathbf{A})$ where $\mathbf{A} \in \{0,1\}^n$ (i.e., if $T_i$ does not appear in any cell of the map for $g(\mathbf{Z})$). To avoid such a situation, one must insist on the *consistency condition* that any atom $T_i$ such that $n_i = 0$ must be forbidden or nullified. This means that the underlying Boolean algebra loses these atoms and hence collapses to a smaller algebra, *i.e.*, to one of its strict sub-algebras. The number of solutions over this new Boolean algebra is

$$N_{conditional} = \prod_{\substack{i=0 \\ n_i \neq 0}}^{K-1} n_i. \tag{15}$$

Note that this number can be 1 (i.e., a unique solution) when $n_i = 1$ for $0 \leq i \leq (K-1)$, but it can grow up to huge values, indeed. Now we introduce a set of parameters $\mathbf{p}_i$ ($0 \leq i \leq (K-1)$, $n_i \neq 0$) to construct an orthonormal set of tags to attach to instances of appearances of the asserted atom $T_i$ in the discriminants $g(\mathbf{A})$ (*i.e.*, in the cells $\mathbf{A} \in \{0,1\}^n$ of the natural map of $g(\mathbf{Z})$). The minimum number of parameters for atom $T_i$ (the length of vector $\mathbf{p}_i$) is given by

$$l(\mathbf{p}_i) = \lceil \log_2 n_i \rceil, \qquad 0 \leq i \leq (K-1), n_i \neq 0. \tag{16}$$

Here, $\lceil x \rceil$ denotes the ceiling of the real number $x$, i.e., the smallest integer greater than or equal to $x$. The parameters $\mathbf{p}_i$ can be used to generate a set of $n_i \leq 2^{l(\mathbf{p}_i)}$ orthonormal tags $\{t_1, t_2 \ldots t_{n_i}\}$, such that

$$t_1 \vee t_2 \vee \ldots \vee t_{n_i} = 1, \tag{17a}$$

$$t_{j_1} \wedge t_{j_2} = 0 \qquad \forall\ j_1, j_2 \in \{1, 2, \ldots, n_i\}. \tag{17b}$$

When $n_i = 2^{l(\mathbf{p}_i)}$ the set of orthonormal tags can be visualized as the products of cells in a Karnaugh map whose map variables are the underlying parameters. If $2^{l(\mathbf{p}_i)-1} < n_i < 2^{l(\mathbf{p}_i)}$, some cells of such a map are merged, and the map reduces to a map-like structure.

When each appearance of an atom $T_i$ is tagged by a particular member of its orthonormal set of tags, an auxiliary function $G(\mathbf{Z}, \mathbf{p}_i)$ ($0 \leq i \leq (K-1)$, $n_i \neq 0$) results. The parametric solution is now given by (Brown, 1970; Rushdi and Amashah, 2011; Rushdi and Ahmad, 2018; Rushdi and Zagzoog, 2019).

$$Z_u = \bigvee_{\{\mathbf{A} \in \{0,1\}^n | A_u = 1\}} G(\mathbf{A}, \mathbf{p}_i).\ 1 \leq u \leq n, (0 \leq i \leq (K-1),\ n_i \neq 0). \tag{18}$$

The total number of parameters used in (18) to construct the tags for all atoms is given by

$$E = \sum_{i=1}^{k} l(\mathbf{p}_i) = \sum_{i=1}^{k} \lceil \log_2(n_i) \rceil. \tag{19}$$

The conventional method is to select the parameter vectors from a shared pool of parameters so as to minimize the number of parameters used, which then becomes

$$E' = \max_i\ l(\mathbf{p}_i) = \max_i\ \lceil \log_2 n_i \rceil = \lceil \log_2 (\max_i n_i) \rceil. \tag{20}$$

However, parameters used must then belong to the underlying Boolean algebra (possibly collapsed due to the consistency condition). We now propose to use independent parameters $\mathbf{p}_i$ for each atom $T_i$ ($0 \leq i \leq K-1$, $n_i \neq 0$). The expressions (18) will not be as compact as they are in the conventional case, but the independent parameters $\mathbf{p}_i$ now belong to the two-valued Boolean algebra $B_2$ (Brown, 2003; Rushdi and Amashah, 2011), a fact that facilitates the generation of all particular solutions as will be seen shortly in the next subsection.

### 3.4. The Subsumptive General Solutions

The disjunctive eliminants (Brown, 1990) or join derivatives (Thayse, 1975) $g_{k+1}(Z_{k+1}, Z_{k+2}, ..., Z_n)$ of the function $g(Z)$ w.r.t. to the set of its first $k$ variables ($k = 0, 1, ..., n$) are defined by

$$g_1(Z_1, Z_2, ..., Z_n) = g(Z_1, Z_2, ..., Z_n), \tag{21a}$$

$$g_k(Z_k, Z_{k+1}, ..., Z_n) = g_{k-1}(1, Z_k, Z_{k+1}, ..., Z_n) \vee g_{k-1}(0, Z_k, Z_{k+1}, ..., Z_n), \ (k = 2, 3, ..., n), \tag{21b}$$

$$g_{n+1} = g_n(1) \vee g_n(0). \tag{21c}$$

The equation (5a) has solutions $Z \in B_2^n$ if it is consistent. This happens if $g_{n+1}$ given by (21c) is 1, in which case the equations $\{g_n = 1, g_{n-1} = 1, ..., g_1 = 1\}$ can be solved in reverse order to yield

$$\overline{g_n}(0) \leq Z_n \leq g_n(1), \tag{22a}$$

$$\overline{g_j}(0, Z_{j+1}, ..., Z_n) \leq Z_j \leq g_j(1, Z_{j+1}, ..., Z_n), \ (j = n-1, ..., 1). \tag{22b}$$

Note that each of the aforementioned equations is a consistency condition for the preceding one (i.e. $g_{k+1} = 1$ is a consistency condition for $g_k = 1$, $k = n, (n-1), ..., 1$). Invocation of these consistency conditions leads to a modification of the solutions (22a) to

$$s_n \leq Z_n \leq t_n, \tag{23a}$$

$$s_k(Z_{k+1}, Z_{k+2}, ..., Z_n) \leq Z_k \leq t_k(Z_{k+1}, Z_{k+2}, ..., Z_n) \quad (k = n-1, n-2, ..., 1), \tag{23b}$$

where

$$\overline{g_n}(0) g_n(1) \leq s_n \leq \overline{g_n}(0), \tag{24a}$$

$$g_n(1) \leq t_n \leq \overline{g_n}(0) \vee g_n(1), \tag{24b}$$

$$\overline{g_k}(0, Z_{k+1}, Z_{k+2}, ..., Z_n) g_k(1, Z_{k+1}, Z_{k+2}, ..., Z_n) \leq s_k \leq \overline{g_k}(0, Z_{k+1}, Z_{k+2}, ..., Z_n), \ (k = n-1, n-2, ..., 1), \tag{24c}$$

$$g_k(1, Z_{k+1}, Z_{k+2}, ..., Z_n) \leq t_k \leq \overline{g_k}(0, Z_{k+1}, Z_{k+2}, ..., Z_n) \vee g_k(1, Z_{k+1}, Z_{k+2}, ..., Z_n), \ (k = n-1, n-2, ..., 1), \tag{24d}$$

This solution is called the double-inequality subsumptive solution. Rushdi (2001; 2004) rewrote it in the following equivalent form, called the incompletely-specified adaptation of the double-inequality subsumptive solution

$$s_n = \overline{g_n}(0) g_n(1) \vee d(\overline{g_n}(0)), \tag{25a}$$

$$t_n = g_n(1) \vee d(\overline{g_n}(0)), \tag{25b}$$

$$s_k = \overline{g_k}(0, Z_{k+1}, Z_{k+2}, ..., Z_n) g_k(1, Z_{k+1}, Z_{k+2}, ..., Z_n) \vee d(\overline{g_k}(0, Z_{k+1}, Z_{k+2}, ..., Z_n)), \ (k = n-1, n-2, ..., 1), \tag{25c}$$

$$t_k = g_k(1, Z_{k+1}, Z_{k+2}, \ldots, Z_n) \lor d(\overline{g_k}(0, Z_{k+1}, Z_{k+2}, \ldots, Z_n)), \quad (k = n-1, n-2, \ldots, 1), \qquad (25d)$$

For convenience, we denote $g_n(0)$ by $g_n(0_n)$, $g_n(1)$ by $g_n(1_n)$, $(0, Z_{k+1}, Z_{k+2}, \ldots, Z_n)$ by $g_k(0_k)$, and $g_k(1, Z_{k+1}, Z_{k+2}, \ldots, Z_n)$ by $g_k(1_k)$. Hence, equations (21) and (25) can be rewritten in the form

$$g_1 = g \qquad (26a)$$

$$g_k = g_{k-1}(0_{k-1}) \lor g_{k-1}(1_{k-1}), \quad (k = 2, 3, \ldots, n+1) \qquad (26b)$$

$$s_k = \overline{g_k}(0_k) \, g_k(1_k) \lor d(\overline{g_k}(0_k)), \quad (k = n, n-1, \ldots, 1), \qquad (26c)$$

$$t_k = g_k(1_k) \lor d(\overline{g_k}(0_k)), \quad (k = n, n-1, \ldots, 1). \qquad (26d)$$

where $d(.)$ stands for the don't-care operator (Reusch, 1975; Rushdi and Albarakati, 2012; 2014). Note that the notation

$$f = g \lor d(h), \qquad (27)$$

means that

$$\{g = 1\} \rightarrow \{f = 1\}, \qquad (28a)$$

$$\{g = 0\} \land \{h = 0\} \rightarrow \{f = 0\} \qquad (28b)$$

Note that (28) keeps silent about the value of $f$ when ($\{g = 0\} \land \{h = 1\}$). The notation (27) is understood (in the case of several variables $X$) to mean

$$f(X) = g(X) \lor (d(X) \land h(X)), \qquad (29)$$

but while the arbitrary function $d(X)$ is clearly understood to be dependent on $X$, *this dependency is hidden* or made implicit to facilitate manipulations. Hiding the dependence of the don't-care symbol $d$ on the pertinent variables is a common practice in the circles of logic design.

### 3.5. Listing of All Particular Solutions

The parametric solutions (18) can be used to generate all particular solutions through the use of an expansion tree. Generally, in the conventional method, this expansion tree is a complete tree that entails the assignment of $2^{K'}$ values to each of $E'$ parameters where $K' \leq K$ is the final number of atoms of the underlying Boolean algebra (possibly after some collapse due to the consistency condition). Each parent node has $2^{K'}$ children nodes and the tree has $E'$ levels beyond its root. Therefore, the tree has $(2^{K'})^{E'} = 2^{K'E'}$ leaves. These leaves constitute the whole set of particular solutions, possibly with repetitions. However, to avoid repetitions, we make sure, right from the first expansion level, to combine any sibling nodes that share the same solution value. With this kind of combining, the tree ceases to be a complete one, and its leaves become exactly the particular solutions, *i.e.*, without repetitions. If we further allow combining cousin (same-level) nodes, the tree is replaced by an acyclic graph that lists particular solution compactly (Rushdi, 2012).

In an alternative method (Rushdi and Ahmad, 2018), a complete version of the expansion tree requires the assignment of binary values {0,1} to each of the $E$ independent parameters. Since the complete binary tree has $E$ levels beyond its root, it has $2^E$ leaves. With merging of sibling nodes of equal solution values, the tree is no longer complete, and its leaves are just the particular solutions without repetitions. The size of the expansion tree in this method is typically less than that in the conventional method since typically $E < K'E'$ (though $E > E'$). However, the true advantage of this alternative method is that *it allows one to avoid the use of an expansion tree (or an expansion acyclic graph) altogether*. The key to this is the observation that the parametric solution (18) can be rewritten as the weighted sum of the atoms $T_i$ that appear in the discriminants $g(A)$ (as expressed in (11)) of the function $g(Z)$, *viz.*

$$\boldsymbol{Z} = \vee_{\substack{i=0 \\ n_i \neq 0}}^{K-1} (\boldsymbol{Co}(T_i) \wedge T_i), \tag{30}$$

where we call the vector $\boldsymbol{Co}(T_i)$ the 'contribution' of the asserted atom $T_i$ and call the conjunction $(\boldsymbol{Co}(T_i) \wedge T_i)$ the 'total contribution' of that atom. We now note that $\boldsymbol{Co}(T_i)$ {or $\boldsymbol{Co}(T_i) \wedge T_i$} has exactly $n_i$ possible values, which can be conveniently listed via the same Karnaugh-map-like structure used in the representation of the associated tags. Therefore, we interpret (30) as a method of conveniently listing all particular solutions as a disjunction of total contributions of asserted atoms $T_i$, where the total contribution is given in all its $n_i$ possibilities. To obtain a specific particular solution, one has simply to pick up arbitrarily one of the possibilities of the total contribution for every atom, and then add the selected total contributions together. The total number of particular solutions obtained this way agrees with that given by (15). The solutions in (30) can alternatively be obtained via the atomic decomposition of pertinent functions and variables (Balamesh and Rushdi, 2019; 2022).

### 3.6. Picking up a Particular Solution of Specific Features

Equation (30) is of a paramount importance, as it provides a listing of a (possibly huge) number of all particular solutions in a compact space. As such, it allows picking up certain solutions enjoying particular desirable features simply by a quick inspection of the aforementioned listing.

## 4. On Algebraic and Map Methods

Algebraic methods constitute the pillars upon which other methods are built and they are inescapable in every implementation detail. Map procedures add nothing to algebraic methods beyond being a *practical way* of organizing them (Rudeanu, 2003). However, such practicality should be not belittled, as it adds *pictorial insight* and *divide-and-conquer strategies* that *facilitate* the manipulations, make them *faster* and *less error prone*, and probably produce more *compact* results (Rushdi and Amashah, 2012). Maps can conveniently tackle laborious manipulations of Boolean functions, which include (a) the unary operation of negation or complementation, which can be applied cell-wise to a function in map form, and (b) any binary operation (such as AND, OR, XOR), which can be implemented cell-wise to two functions both in map form. Moreover, map folding (Rushdi, 1986; Rushdi and Rushdi, 2018) directly implements the operations of meet and join differentiation of a Boolean function with respect to one or several of its variables (also called conjunctive or disjunctive elimination of these variables).

Map methods are not confined to the use of the classical Karnaugh map (CKM) but they overlap with algebraic methods through the use of the variable-entered Karnaugh map (VEKM), which is semi algebraic in nature (Rushdi and Amashah, 2011). A Boolean function of n variables has $2^n$ VEKM representations (depending on the choice of map and entered variables) ranging from a CKM (of n map variables and 0 entered variables), to a purely-algebraic expression (of 0 map variables and n entered variables). This means clearly that a purely-algebraic expression is *a special case of* a VEKM representation. Of course, the VEKM cannot handle infinite Boolean algebras, but it serves as a *natural map* for finite 'big' Boolean algebras (Brown, 1990). There is no point of claiming that algebraic methods are superior to VEKM methods, since the latter include the former as special cases. On the contrary, VEKM methods, being semi algebraic in nature, can always take full advantage of the results provided by the algebraic theory. They have a better control on the minimality of the pertinent function representation.

Specifically, Rushdi and Amashah (2012) studied the purely-algebraic method of Rudeanu (2003) and showed that it derives its simplicity from sacrificing a degree of freedom offered by the double-inequality subsumptive solution; thereby employing a special completely-specified instant of that solution. Hence, it secures *local minimality* over a set of chosen coefficients and not over the more basic set of pertinent variables and generators. Therefore, this purely-algebraic method is not always as efficient as a general VEKM method, which seeks *global minimality* based on an incompletely-specified adaptation of the double-inequality subsumptive solution. This means that VEKM solutions are at least as good as purely-algebraic ones, but are occasionally better.

## 5. An Illustrative Example

The problem studied in this section is taken from an old text on Boolean algebra (Goodstein, 1963) that supplied a general parametric "solution" *via* a non-constructive theorem-proof technique. However, this alleged solution fails to generally satisfy the equation it is intended to solve since it does not recognize the need for a consistency condition. Consider the Boolean function

$$f(X, Y) = c\,(a \vee X)\,(b \vee Y), \tag{31}$$

where $f: B_{256}^2 \to B_{256}$, and $B_{256} = FB(a, b, c)$. A solution of the equation $\{f = 0\}$ expresses the dependent variables X and Y in terms of the independent " variables" a, b and c which are treated herein as generators of the underlying Boolean algebra. We complement the function $f$ to obtain $g = \bar{f}$ so as to solve the equivalent equation

$$g(X, Y) = \bar{c} \vee a\bar{X} \vee \bar{b}\bar{Y} = 1, \tag{32}$$

where $g: B_{256}^2 \to B_{256}$. In the next three subsections, we offer three solutions of (32). These are a conventional parametric solution obtained directly, a conventional parametric solution obtained via a subsumptive one, and a permutative additive parametric solution.

### 5.1. A Conventional Parametric Solution Obtained Directly

The Boole-Shannon expansion of *g w. r. t.* its two arguments X and Y is

$$\begin{aligned}g(X, Y) &= (\bar{a} \vee \bar{b} \vee \bar{c})\bar{X}\,\bar{Y} \vee (\bar{a} \vee \bar{c})\bar{X}\,Y \vee (\bar{b} \vee \bar{c})X\,\bar{Y} \vee (\bar{c})X\,Y \\ &= (\bar{a}bc \vee \bar{a}\bar{b}c \vee a\bar{b}c \vee \bar{c})\bar{X}\,\bar{Y} \vee (\bar{a}bc \vee \bar{a}\bar{b}c \vee \bar{c})\bar{X}\,Y \vee (a\bar{b}c \vee \bar{a}\bar{b}c \vee \bar{c})X\,\bar{Y} \vee (\bar{c})X\,Y,\end{aligned} \tag{33}$$

and hence its natural map (variable-entered Karnaugh map (VEKM)) is as shown in Fig. 3. Each of the entries of this map is a function of the "entered variables" or generators a, b and c and is a disjunction of some of the eight atoms of FB(a, b, c). The numbers of appearances of these eight atoms in the cells of the map of Fig. 3 are 4,3,4,2,4,0,4, and 2, which immediately shows that:

1. The atom abc does not appear at all in any of the cells of the map in Fig. 3. This means that this atom must be nullified, *i.e.*, the consistency condition of the equation (32) is

$$abc = 0. \tag{34}$$

When the Boolean algebra $B_{256} = FB(a, b, c)$ loses its abc atom, it *collapses* into a sub-algebra of 7 atoms only, *i.e.*, it collapses to $B_{128}$. The sizes of $B_{256}$ and $B_{128}$ are too large to be amenable to visualization here.

However, to give the reader a glimpse of the meaning of algebra collapse, we have already presented in Fig. 1 a hypercube lattice representing $B_{16}$, and then represented its collapse to $B_8$ when it lost the $ab$ atom.

2. The number of particular solutions of (32) is the product of the numbers of appearances of asserted atoms in Fig. 3, namely

$$N_{particular} = 2^2 * 3^1 * 4^4 = 3072. \tag{35}$$

The minimum number of parameters $k$ needed for a parametric solution of (32) is

$$k = \lceil \log_2 4 \rceil = 2. \tag{36}$$

In fact, this number is expected to be less than or equal to the number of variables involved (Brown, 1990), which is $n = 2$ herein.

To facilitate obtaining a parametric solution with a minimum number of parameters, we note that each of four atoms $\bar{a}\bar{b}\bar{c}$, $\bar{a}b\bar{c}$, $a\bar{b}c$, and $ab\bar{c}$ is omnipresent in the map of Fig. 3, and therefore we kept them combined into $\bar{c}$. Note that such a combination is permissible since it is equivalent to using the same set of orthonormal tags individually in the same way with instances of each of these atoms or collectively with their total disjunction $\bar{c}$. In Fig. 4, we construct an auxiliary function $G_1(X, Y; a, b, c; u, v)$, where we attach tags from the orthonormal set $\{\bar{u}\bar{v}, \bar{u}v, u\bar{v}, uv\}$ to the term $\bar{c}$ (that has 4 appearances), attach tags from the orthonormal set $\{(\bar{u}\bar{v} \vee uv), \bar{u}v, u\bar{v}\}$ to the atom $\bar{a}\bar{b}c$ (that has 3 appearances), and attach tags from the orthonormal set $\{(\bar{u} \vee v), u\bar{v}\}$ to atom $a\bar{b}c$ (that has 2 appearances). We have chosen the orthonormal sets above with an eye on getting the most compact solution. In fact, we do not care about how cumbersome the entries in the $\bar{X}\bar{Y}$-cell are, since they do not affect the final solution. However, we have only a single tag per each of the other three cells, namely tag $\bar{u}v$ in the $\bar{X}Y$-cell, tag $u\bar{v}$ in the $X\bar{Y}$-cell, and tag $uv$ in the $XY$-cell. We might add the nullified atom abc don't-care in each of the cells of Fig. 4. Our final solution is

$$X = (\bar{a}\bar{b}c \vee a\bar{b}c \vee \bar{c})u\bar{v} \vee \bar{c}uv \vee d(abc) \tag{37a}$$
$$Y = (\bar{a}\bar{b}c \vee a\bar{b}c \vee \bar{c})\bar{u}v \vee \bar{c}uv \vee d(abc). \tag{37b}$$

These formulas might be simplified by ignoring the don't-care parts and involving the reflection law to obtain

$$X = u(\bar{b}c\bar{v} \vee \bar{c}(\bar{v} \vee v)) = u(\bar{b}\bar{v} \vee \bar{c}) \tag{38a}$$
$$Y = v(\bar{b}c\bar{u} \vee \bar{c}(\bar{u} \vee u)) = v(\bar{a}\bar{u} \vee \bar{c}) \tag{38b}$$

Substitution of the solution (38) in (32) yields

$$f(X, Y) = c(a \vee u\bar{b}\bar{v} \vee u\bar{c})(b \vee v\bar{a}\bar{u} \vee v\bar{c}) = abc = 0, \tag{39}$$

while its substitution in (33) yields

$$g(X, Y) = \bar{c} \vee \bar{a}(\bar{u} \vee (b \vee v)c) \vee \bar{b}(\bar{v} \vee (a \vee u)c)$$
$$= \bar{c} \vee \bar{a}\bar{u} \vee \bar{a}b \vee \bar{a}v \vee \bar{b}\bar{v} \vee a\bar{b} \vee \bar{b}u(\bar{u} \vee (b \vee v)c) = \bar{a} \vee \bar{b} \vee \bar{c} = 1, \tag{40}$$

where the consensus $\bar{a}\bar{b}$ of $\bar{a}\bar{u}$ and $\bar{b}u$ is added to the second line in (40), and then combined with $(\bar{a}b \vee a\bar{b})$ to produce $(\bar{a} \vee \bar{b})$ which then absorbs $(\bar{a}\bar{u} \vee \bar{a}v \vee \bar{b}\bar{v} \vee \bar{b}u)$. The results of the aforementioned substitution verify the solution and partially explains why the consistency condition is needed. We further employed $(\bar{a} \vee \bar{b} \vee \bar{c}) = 1$, thanks to the consistency condition in (34), since this results by complementing both sides in (34). The algebraic manipulations above could have been much simplified if we resorted to map minimization techniques (Rushdi, 1987).

## 5.2. A Conventional Parametric Solution obtained *via* a Subsumptive solution

In this subsection, we utilize Variable-Entered Karnaugh Maps (VEKMs) to derive a general subsumptive solution for (29) in the most compact form. Figure 5 presents the VEKMs used to obtain such a solution according to the method of Rushdi (2001; 2004). The final result obtained is

$$0 \leq Y \leq \bar{c} \vee \bar{a}\, \overline{X} \tag{41a}$$
$$0 \leq X \leq \bar{b} \vee \bar{c}, \tag{41b}$$

subject to the consistency solution

$$\bar{a} \vee \bar{b} \vee \bar{c} = 1. \tag{41c}$$

We can convert the subsumptive solution (41) into a parametric one, namely

$$X = u(\bar{b} \vee \bar{c}) \tag{42a}$$
$$Y = v(\bar{c} \vee \bar{a}\overline{X}) = v(\bar{c} \vee \bar{a}\bar{u} \vee \bar{a}bc) = v(\bar{c} \vee \bar{a}\bar{u} \vee \bar{a}b). \tag{42b}$$

Substituting this solution into (31), one obtains

$$f(X, Y) = c(a \vee u\bar{b} \vee u\bar{c})(b \vee v\bar{c} \vee v\bar{a}\bar{u} \vee v\bar{a}b) = abc = 0. \tag{43}$$

This solution is not symmetric like the one in (38). Hence, it can be used to generate a third solution, *viz.*,

$$Y = v(\bar{a} \vee \bar{c}) \tag{44a}$$
$$X = u(\bar{c} \vee \bar{b}\overline{Y}) = u(\bar{c} \vee \bar{b}\bar{v} \vee a\bar{b}) = v(\bar{c} \vee \bar{a}\bar{u} \vee \bar{a}b). \tag{44b}$$

Though the solutions (42) and (44) are not symmetric like the one in (38), they enjoy the advantage that in each of them one variable is dependent on a single parameter rather than on the two parameters.

## 5.3. A Permutative Additive Parametric Solution

Despite the elegance, compactness, and symmetry of the solutions in (38), (42) or (44), they are not readily useful for producing a list of all particular solutions, since each of the two parameters $u$ and $v$ should be assigned an independent value that equals a specific element in $B_{128}$. The expansion tree used for this purpose should explore all $128 \times 128 = 16384$ combinations of $(u, v)$ values, and will finally settle on 3072 solutions. Our alternative method to avoid the use of such an expansion tree is to use independent parameters for each individual atom, as shown in Fig. 6. The number of parameters used increases dramatically from 2 to 12, and though a detailed algebraic solution will be cumbersome when compared with the earlier solutions in (38), (42) or (44), it is nevertheless a permutative additive formula of an atomic-decomposition nature, and it lists all 3072 particular solutions of equation (32) compactly. This formula is given concisely in Fig. 7. To obtain the value for the vector $[X\, Y]^T$, one chooses any of the possible values associated with each atom. Two examples of the particular solutions (subject to the condition $abc = 0$) are

$$\begin{bmatrix} X \\ Y \end{bmatrix} = \bar{a}\bar{b}\bar{c}\begin{bmatrix} 0 \\ 0 \end{bmatrix} \vee \bar{a}\bar{b}c\begin{bmatrix} 0 \\ 0 \end{bmatrix} \vee \bar{a}b\bar{c}\begin{bmatrix} 0 \\ 0 \end{bmatrix} \vee \bar{a}bc\begin{bmatrix} 0 \\ 0 \end{bmatrix} \vee a\bar{b}\bar{c}\begin{bmatrix} 0 \\ 0 \end{bmatrix} \vee a\bar{b}c\begin{bmatrix} 0 \\ 0 \end{bmatrix} \vee ab\bar{c}\begin{bmatrix} 0 \\ 0 \end{bmatrix} \vee abc\begin{bmatrix} 0 \\ 0 \end{bmatrix} = \begin{bmatrix} 0 \\ 0 \end{bmatrix} \tag{45}$$

$$\begin{bmatrix} X \\ Y \end{bmatrix} = \bar{a}\bar{b}\bar{c}\begin{bmatrix} 0 \\ 1 \end{bmatrix} \vee \bar{a}\bar{b}c\begin{bmatrix} 0 \\ 1 \end{bmatrix} \vee \bar{a}b\bar{c}\begin{bmatrix} 0 \\ 1 \end{bmatrix} \vee \bar{a}bc\begin{bmatrix} 0 \\ 1 \end{bmatrix} \vee a\bar{b}\bar{c}\begin{bmatrix} 1 \\ 0 \end{bmatrix} \vee a\bar{b}c\begin{bmatrix} 1 \\ 0 \end{bmatrix} \vee ab\bar{c}\begin{bmatrix} 1 \\ 0 \end{bmatrix} \vee abc\begin{bmatrix} 1 \\ 0 \end{bmatrix} = \begin{bmatrix} a \\ \bar{a} \end{bmatrix}. \tag{46}$$

Note that any of the particular solutions if substituted in (31) reduces it to $\{abc = 0\}$, and if substituted in (32) reduces it to $\{\bar{a} \vee \bar{b} \vee \bar{c} = 1\}$.

## 6. Conclusions

This paper is a review of the contemporary status of the two-century old topic of 'big' Boolean-equation solving. The paper stresses that this topic has matured enough to find many significant, diverse, and beneficial applications, notably in digital design and cryptography. The paper gives full descriptions and demonstrations of methods to construct parametric or subsumptive general solutions and to compactly list particular solutions of 'big' Boolean equations. As an offshoot, the paper explains the necessity of imposing a consistency condition and the possible impact of such a condition on collapsing the underlying Boolean algebra to a strictly smaller sub-algebra. As a result, the paper sets the stage for future full utilization of the mathematics of 'big' Boolean algebras and 'big' Boolean equation-solving.

|  | X |
|---|---|
| $\bar{a}\bar{b}c \lor \bar{a}bc \lor a\bar{b}c \lor \bar{c}$ | $\bar{a}bc \lor a\bar{b}c \lor \bar{c}$ |
| Y $\bar{a}\bar{b}c \lor \bar{a}bc \lor \bar{c}$ | $\bar{c}$ |

$$g(X, Y)$$

Fig. 3. The natural map of $g(X, Y)$, with the four atoms $\bar{a}\bar{b}\bar{c}, \bar{a}b\bar{c}, a\bar{b}\bar{c}$ and $ab\bar{c}$ combined as $\bar{c}$. Such a combination is equivalent to using the same set of orthonormal tags $\{\bar{u}\bar{v}, \bar{u}v, u\bar{v}, uv\}$ individually in the same way with instances of each of these atoms or collectively with their total disjunction $\bar{c}$.

|  | X |
|---|---|
| $\bar{a}\bar{b}c\,(\bar{u}\bar{v} \lor uv) \lor$<br>$\bar{a}bc\,(u \lor \bar{v}) \lor$<br>$a\bar{b}c\,(\bar{u} \lor v) \lor$<br>$\bar{c}\,(\bar{u}\bar{v}) \lor$<br>$d(abc)$ | $\bar{a}bc\,(u\bar{v}) \lor$<br>$a\bar{b}c\,(u\bar{v}) \lor$<br>$\bar{c}\,(u\bar{v}) \lor$<br>$d(abc)$ |
| Y $\bar{a}\bar{b}c\,(\bar{u}v) \lor$<br>$\bar{a}bc\,(\bar{u}v) \lor$<br>$\bar{c}\,(\bar{u}v) \lor$<br>$d(abc)$ | $\bar{c}\,(uv) \lor$<br>$d(abc)$ |

$$G_1(X, Y; a, b, c; u, v)$$

Fig. 4. The auxiliary function $G_1$ with instances of each atom tagged by members of an orthonormal set. Common parameters are used for different atoms.

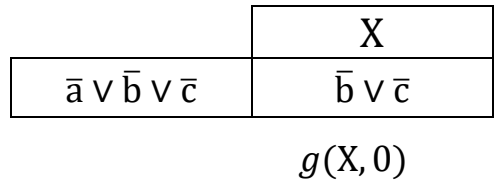

$$g(X, 0)$$

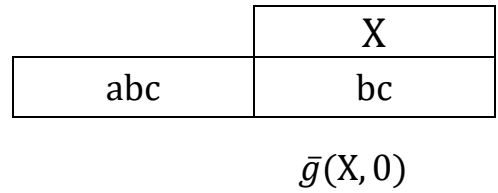

$$\bar{g}(X, 0)$$

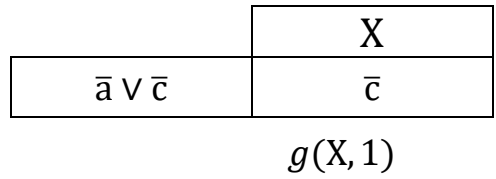

$$g(X, 1)$$

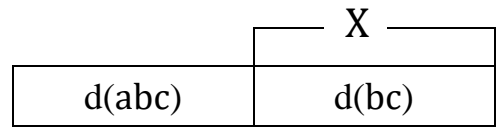

$$s_Y = \bar{g}(X, 0)g(X, 1) \vee d(\bar{g}(X, 0)) = 0$$

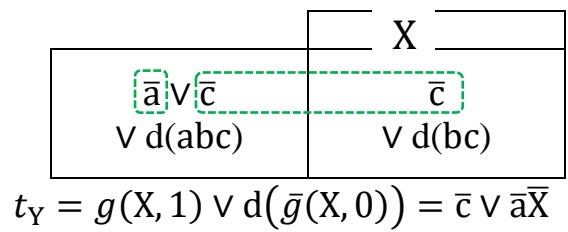

$$t_Y = g(X, 1) \vee d(\bar{g}(X, 0)) = \bar{c} \vee \bar{a}\bar{X}$$

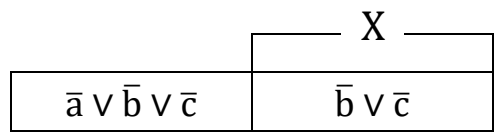

$$g_1(X) = g(X, 0) \vee g(X, 1) = 0$$

$$\boxed{\bar{a} \lor \bar{b} \lor \bar{c}}$$

$$g_1(0)$$

$$\boxed{abc}$$

$$\overline{g_1}(0)$$

$$\boxed{\bar{b} \lor \bar{c}}$$

$$g_1(1)$$

$$\boxed{d(abc)}$$

$$s_X = \overline{g_1}(0)\, g_1(1) \lor d(\overline{g_1}(0)) = 0$$

$$\boxed{\begin{array}{c} \bar{b} \lor \bar{c} \\ \lor\, d(abc) \end{array}}$$

$$t_X = g_1(1) \lor d(\overline{g_1}(0)) = \bar{b} \lor \bar{c}$$

$$\boxed{\bar{a} \lor \bar{b} \lor \bar{c}}$$

$$g_0 = g_1(0) \lor g_1(1) = 1$$

Fig. 5. Various VEKMs used in the derivation of the most compact subsumptive solution of (32).

|   | X |
|---|---|
| $\bar{a}\bar{b}\bar{c}\ \bar{p}_1\bar{p}_2\ \vee$<br>$\bar{a}\bar{b}c\ p_3\bar{p}_4\ \vee$<br>$\bar{a}b\bar{c}\ \bar{p}_5 p_6\ \vee$<br>$\bar{a}bc\ \bar{p}_7\ \vee$<br>$a\bar{b}\bar{c}\ \bar{p}_8\bar{p}_9\ \vee$<br>$a\bar{b}c\ \bar{p}_{10}\ \vee$<br>$ab\bar{c}\ \bar{p}_{11}\bar{p}_{12}\ \vee$<br>$d(abc)$ | $\bar{a}\bar{b}\bar{c}\ p_1\bar{p}_2\ \vee$<br>$\bar{a}\bar{b}c\ \bar{p}_3\bar{p}_4\ \vee$<br>$\bar{a}b\bar{c}\ p_5\bar{p}_6\ \vee$<br><br>$a\bar{b}\bar{c}\ p_8\bar{p}_9\ \vee$<br>$a\bar{b}c\ p_{10}\ \vee$<br>$ab\bar{c}\ p_{11}\bar{p}_{12}\ \vee$<br>$d(abc)$ |
| $\bar{a}\bar{b}\bar{c}\ \bar{p}_1 p_2\ \vee$<br>$\bar{a}\bar{b}c\ p_4\ \vee$<br>$\bar{a}b\bar{c}\ \bar{p}_5 p_6\ \vee$<br>$\bar{a}bc\ p_7\ \vee$<br>$a\bar{b}\bar{c}\ \bar{p}_8 p_9\ \vee$<br><br>$ab\bar{c}\ \bar{p}_{11} p_{12}\ \vee$<br>$d(abc)$ | $\bar{a}\bar{b}\bar{c}\ p_1 p_2\ \vee$<br><br>$\bar{a}b\bar{c}\ p_5 p_6\ \vee$<br><br>$a\bar{b}\bar{c}\ p_8 p_9\ \vee$<br><br>$ab\bar{c}\ p_{11} p_{12}\ \vee$<br>$d(abc)$ |

(Y labels the bottom row)

$$G_2(X, Y;\ a, b, c;\ p)$$

Fig. 6. The auxiliary function $G_2$ with instances of each atom tagged by members of an orthonormal set. Independent parameters are used for different atoms, and hence combining the four $\bar{c}$ atoms does not work any more.

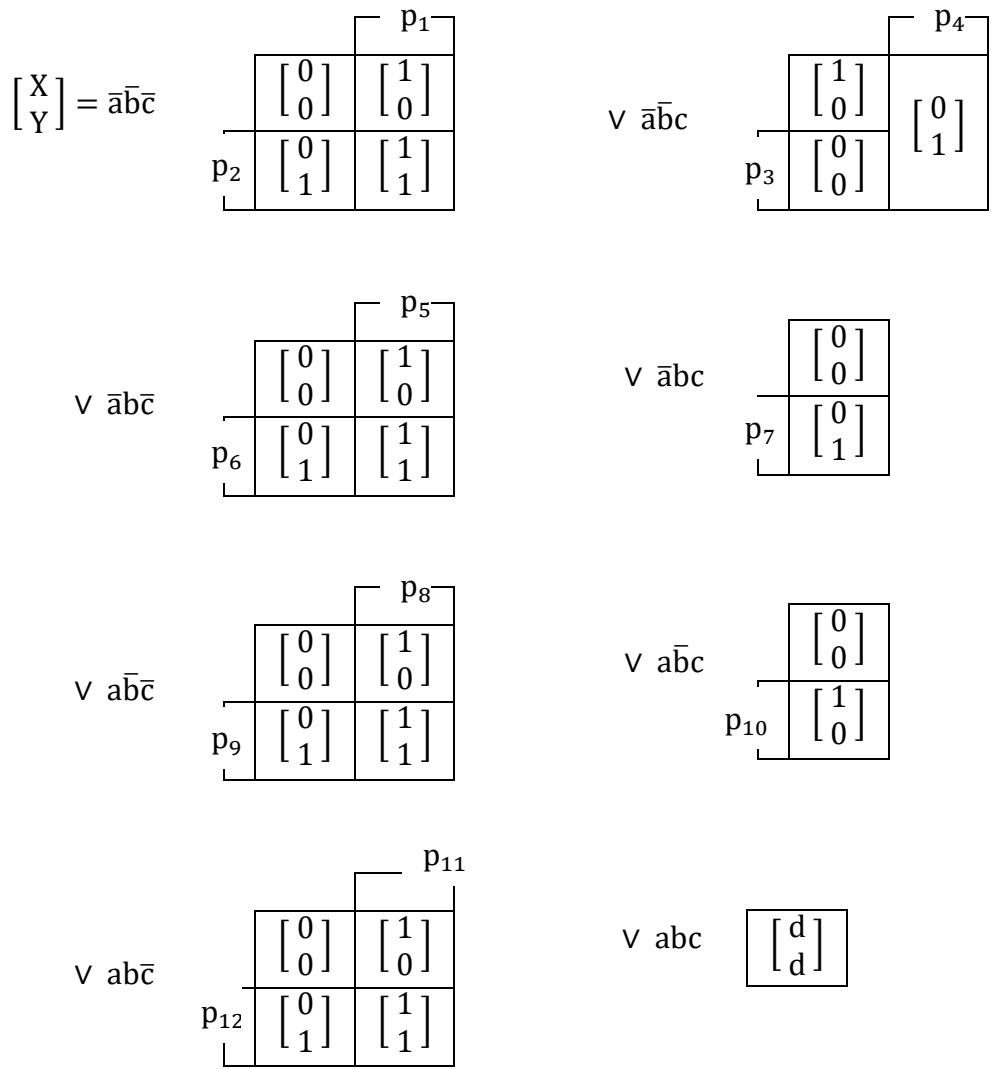

Fig. 7. A permutative additive formula listing all the 3072 particular solutions of Equation (32).

**Acknowledgment**
A brief version of this paper has been presented as an **invited keynote talk** in a plenary session of the First International Conference on Mathematical Methods and Techniques in Engineering and Sciences (ICMMTES 2022), held on December 9-10, 2022 at the Graphic Era Deemed to be University | Graphic Era Hill University, Dehradun, Uttarakhand, INDIA.